\documentclass[%
 reprint,
superscriptaddress,
 amsmath,amssymb,
 aps,
prapplied,
]{revtex4-2}
\usepackage{graphicx}
\usepackage{dcolumn}
\usepackage{bm}
\usepackage{hyperref}
\usepackage{xcolor}

\usepackage{comment}
\usepackage[separate-uncertainty=true]{siunitx}
\usepackage[siunitx]{circuitikz}

\begin{document}

\preprint{APS/123-QED}

\title{Enhancing Membrane-Based Scanning Force Microscopy Through an Optical Cavity}

\author{Thomas Gisler}
\thanks{These authors contributed equally.\\ Corresponding author: \href{mailto:vdumont@phys.ethz.ch}{vdumont@phys.ethz.ch}} 
\affiliation{Laboratory for Solid State Physics, ETH Zürich, CH-8093 Zürich, Switzerland}

\author{David Hälg}
\thanks{These authors contributed equally.\\ Corresponding author: \href{mailto:vdumont@phys.ethz.ch}{vdumont@phys.ethz.ch}}
\affiliation{Laboratory for Solid State Physics, ETH Zürich, CH-8093 Zürich, Switzerland}

\author{Vincent Dumont}
\thanks{These authors contributed equally.\\ Corresponding author: \href{mailto:vdumont@phys.ethz.ch}{vdumont@phys.ethz.ch}} 
\affiliation{Laboratory for Solid State Physics, ETH Zürich, CH-8093 Zürich, Switzerland}

\author{Shobhna Misra}
\affiliation{Laboratory for Solid State Physics, ETH Zürich, CH-8093 Zürich, Switzerland}

\author{Letizia Catalini}
\affiliation{Laboratory for Solid State Physics, ETH Zürich, CH-8093 Zürich, Switzerland}

\author{Eric C. Langman}
\affiliation{Niels Bohr Institute, University of Copenhagen, Copenhagen 2100, Denmark}
\affiliation{Center for Hybrid Quantum Networks, Niels Bohr Institute, University of Copenhagen, Copenhagen 2100, Denmark}

\author{Albert Schliesser}
\affiliation{Niels Bohr Institute, University of Copenhagen, Copenhagen 2100, Denmark}
\affiliation{Center for Hybrid Quantum Networks, Niels Bohr Institute, University of Copenhagen, Copenhagen 2100, Denmark}

\author{Christian L. Degen}
\affiliation{Laboratory for Solid State Physics, ETH Zürich, CH-8093 Zürich, Switzerland}

\author{Alexander Eichler}
\affiliation{Laboratory for Solid State Physics, ETH Zürich, CH-8093 Zürich, Switzerland}

\date{\today}

\begin{abstract}
The new generation of strained silicon nitride resonators harbors great promise for scanning force microscopy, especially when combined with the extensive toolbox of cavity optomechanics. However, accessing a mechanical resonator inside an optical cavity with a scanning tip is challenging. Here, we experimentally demonstrate a cavity-based scanning force microscope based on a silicon nitride membrane sensor. We overcome geometric constraints by making use of the extended nature of the mechanical resonator normal modes, which allows us to spatially separate the scanning and readout sites of the membrane. Our microscope is geared towards low-temperature applications in the zeptonewton regime, such as nanoscale nuclear spin detection and imaging.
\end{abstract}

\maketitle

\section{Introduction}

Resonators with ultrahigh quality factors have become prominent nanomechanical platforms~\cite{Bachtold_RMP2022}. Dissipation dilution through in-plane strain in silicon nitride~\cite{schmid_2011} can be combined with geometrical patterning to minimize clamping loss through soft clamping~\cite{Tsaturyan2017Ultracoherent,Ghadimi2018Elastic}, perimeter modes~\cite{Bereyhi_2022}, hierarchical clamping~\cite{fedorov2020fractal, bereyhi2022hierarchical}, and spiderweb designs~\cite{shin2022spiderweb, hoj2021ultra}. Such mechanical resonators attain quality factors greater than $10^9$~\cite{Rossi2018Measurement,gisler2022soft,seis2022ground,Bereyhi_2022,bereyhi2022hierarchical,shin2022spiderweb} at cryogenic temperatures, resulting in exceedingly long coherence times and thermally limited force noise below $\SI{100}{zN/Hz^{1/2}}$~\cite{eichler2022ultra}.

Scanning force microscopy is one area where the mechanical properties of such resonators could be exploited. Indeed, the outstanding force sensitivity of these devices makes them ideal candidates for applications such as magnetic force microscopy~\cite{christensen20242024} and nanoscale magnetic resonance imaging (NanoMRI)~\cite{budakian2023roadmap}, whose operational scope depends crucially on the sensor performance. This new generation of nanomechanical sensors could potentially allow three-dimensional NanoMRI with near-atomic resolution~\cite{eichler2022ultra,visani2023near}, which would offer significant benefits for life science research and material science~\cite{budakian2023roadmap}. An important challenge of state-of-the-art scanning force demonstrations with ultracoherent membranes~\cite{Scozzaro_2016,halg2021membrane} and trampoline resonators~\cite{fischer2019spin} is optical detection noise generated by the interferometric readout. Currently, the excellent resonator characteristics do not yet yield a benefit at cryogenic temperatures, because the  force noise budget is dominated by the readout noise, especially at larger bandwidths.

One avenue towards reducing the readout noise is  positioning the mechanical element within a high-finesse cavity~\cite{Thompson2008Strong, Jayich2008Dispersive}. Doing so additionally allows for the diverse cavity optomechanics toolbox to be employed~\cite{Markuseditor2014Cavity, Aspelmeyer2014Cavity, bowen2015quantum}. For instance, it becomes possible to simultaneously sideband cool multiple mechanical modes in the `fast-cavity' limit~\cite{Aspelmeyer2014Cavity} and to use the correlations in the cavity light to make force measurements near~\cite{lahaye2004approaching,Rossi2018Measurement,Schreppler2014Quantum} or even below the standard quantum limit~\cite{Mason2019Continuous}. Furthermore, a cavity can reduce optical heating (see Appendix~\ref{app:laser_heating}), which is especially important for cryogenic operation~\cite{planz2023membrane}.

While the lower readout noise and the optomechanics toolbox are important assets, incorporating a high-finesse cavity in such a scanning force instrument poses significant challenges. For instance, approaching a scanning tip close to a membrane sandwiched between two mirrors is difficult due to space constraints. So far, optomechanical scanning force microscopes have only been realized using specialized cantilever designs~\cite{guha2020force, fogliano2021mapping, srinivasan2011optomechanical}. The use of silicon nitride membrane sensors, whose large stiffness will help reduce non-contact friction~\cite{heritier2021spatial, kuehn2006dielectric} and `pinning'~\cite{krass2022force} of the mechanical sensor, would allow for a significant leap forward. Furthermore, once cooled, this platform should be quantum-limited (see Appendix~\ref{App:Sec-Prospects}) and lead to, amongst other things, quantum-limited atomic force microscope~\cite{milburn1994quantum}.

In this paper, we present a high-finesse membrane-in-the-middle cavity scanning force experiment. Opposite to a previous membrane-based scanning microscopy demonstration~\cite{halg2021membrane}, this setup reaches the thermal-noise limit of the membrane sensor over a larger signal bandwidth, offering a much better overall noise budget. Specifically, we demonstrate a readout noise of \SI{1}{\femto\meter\per\hertz^{1/2}} and a thermally limited force noise of \SI{180}{\atto\newton\per\hertz^{1/2}} at room temperature over a bandwidth of \SI{80}{Hz}. This improvement is essential in the context of force-detected NanoMRI~\cite{sidles1991,rugar_2004single,Degen_2009,poggio2010force,vinante2011magnetic,Nichol_2013,grob_magnetic_2019,haas2022nuclear}, as it will enable higher spatial resolution and faster scans. Combined with state-of-the-art membranes~\cite{Rossi2018Measurement} and nanoscale dynamic nuclear polarization~\cite{tabatabaei2024large}, this setup is ideally positioned to help realize recent proposals for nuclear spin detection and imaging~\cite{Kosata_2020,visani2023near}.

\section{Cavity-Based Scanning Force Sensing Apparatus}

\begin{figure}[t]
	\centering
	\includegraphics[width=1\columnwidth]{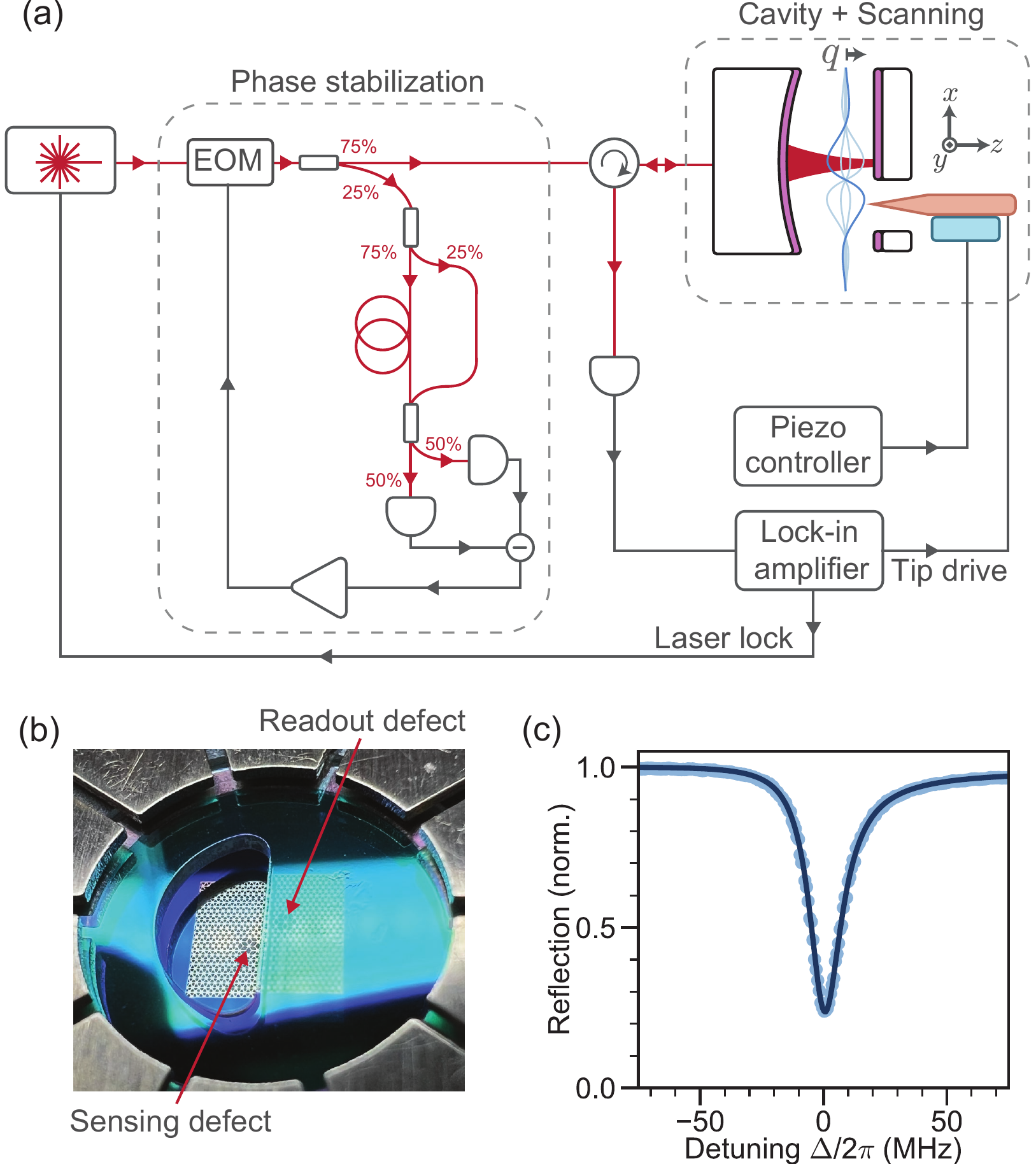}
	\caption{Cavity-based membrane scanning force sensor. (a)~Experimental schematic. The laser phase noise is reduced through a phase stabilization loop before being sent to the cavity-based membrane scanning force sensor. The reflected light amplitude (and phase) of the locked cavity provides a record of the membrane vibration amplitude and frequency. (b)~Photograph of the double-defect patterned membrane placed in between two mirrors. The `readout defect' is inside the membrane-in-the-middle cavity, while the `sensing defect' can be accessed with a scanning tip through the `D'-shape cutout. (c)~Reflected light from the cavity as a function of laser detuning $\Delta$. The resonance dip can be fitted to a Fano lineshape~\cite{Janitz2015Fabry}, yielding cavity power decay rate $\kappa/2\pi = \SI{15.9(1)}{MHz}$, which corresponds to finesse $\mathcal{F} =\SI{3.92(3)e3}{}$.  }
	\label{Fig:1} 
\end{figure}

We present the schematic of our cavity-based membrane force scanner in Fig.~\ref{Fig:1}(a). Its sensor, shown in Fig.~\ref{Fig:1}(b), consists of a 20-nm-thick high-stress silicon nitride  membrane patterned with a phononic shield~\cite{Tsaturyan2017Ultracoherent} with two mechanical `defects' near its center~\cite{catalini2020soft,halg2021membrane}. These defects are physically close enough ($\sim \SI{1.2}{mm}$) for their individual modes to hybridize into coupled symmetric and anti-symmetric normal modes. For our experiment, we use the anti-symmetric mode with a resonance frequency $\Omega_{\mathrm{m}}/2\pi \approx \SI{1.424}{MHz}$, an effective mass $m_\mathrm{eff} \approx\SI{20}{ng}$ calculated from a COMSOL simulation, and a quality factor $Q_\mathrm{m} = \SI{88.5(2)e6}{}$ extracted from ringdown measurements.

The membrane is placed between two mirrors, forming a membrane-in-the-middle optomechanical cavity~\cite{Thompson2008Strong, Jayich2008Dispersive}. A laser (Toptica CTL1550) with tuneable wavelength (nominally $\lambda = \SIrange[range-phrase={-}]{1510}{1630}{nm}$) passes through an electro-optic modulator (EOM, iXBlue MPZ-LN-10), which we use to actively stabilize the laser phase via feedback. This phase stabilization works by sending 25\% (Koheron FOSPL12-7525) of the light to a self-homodyne phase-detection scheme \cite{parniak2021high}. The light is split into two paths by a second beamsplitter (Koheron  FOSPL12-7525), with 75\% of the light going through an optical delay line (i.e., a \SI{43}{m}-long optical fiber)~\footnote{The optical delay line length can also be controlled with our homemade fiber stretcher. The \SI{43}{m}-optical fiber is wrapped around two 3D-printed half-cylinders, which are glued to two piezolectric elements (Thorlabs PA4HKW). By applying a voltage to the piezos, the distance between the two half-cylinders increases, slightly stretching the fiber and increasing the optical path length. This allows to read out an arbitrary quadrature of the light, and stabilize the optical path length. Note that we also control the polarization (Thorlabs MPC320) and amplitude (Thorlabs V1550A) of the light in this arm, so it matches the light of the other path before mixing on the 50:50 beamsplitter.}, before being recombined with the other (non-delayed) path with 25\% of the light on a 50:50 beamsplitter (Thorlabs TN1550R5A2). The light at the two output ports of the beamsplitter is measured with a balanced photodetector (Koheron PD10B-3-DC), creating a homodyne signal which is amplified and filtered (RedPitaya STEMlab 125-14) before being fed back to the EOM. With this phase stabilization scheme, we reduce our laser frequency noise from \SI{5.6}{Hz/Hz^{1/2}} to \SI{0.33}{Hz/Hz^{1/2}} over a frequency range $\SIrange[range-phrase={-}]{1.42}{1.44}{MHz}$ comprising the mechanical mode, see Appendix~\ref{app:laser_phase_noise}. 

The phase-stabilized laser is then sent to a circulator (Koheron  FOCIR1550-A) before reaching the scanning cavity. The optical signal reflected from the cavity is then detected by a photodiode in direct detection (Thorlabs FGA01FC) and a transimpedance amplifier (Femto DHPCA-100), and the signal is measured by a lock-in amplifier (Zurich Instruments HF2LI or MFLI) providing a record of the membrane motion. The reflected signal is also used to lock the laser frequency to a fixed cavity detuning with respect to the cavity resonance frequency using a side-of-fringe lock, via the laser's internal locking scheme (Toptica DLC Pro).

The extended mechanical normal modes on the membrane allow us to spatially separate the sample position, which has to be accessed by the scanning tip, from the optical cavity. The membrane acts as the mechanical sensor in an inverted atomic force microscopy (AFM) geometry, with samples placed on the membrane surface and approached by a  non mechanically compliant scanning tip~\cite{halg2021membrane}. Here, the membrane is placed between two mirrors, of which the back one consists of a Bragg mirror evaporated  on a flat sapphire substrate with specified transmission $|t|^2<10^{-4}$ (Layertec~\footnote{These mirrors are made by first using a  \SI{0.7}{mm}-thick polished sapphire substrate, in which we laser-cut a `D'-hole. We repolish them for 15 mins (Logitech PM5 autolap) and clean them afterwards in an acetone bath overnight, without sonication. We finally send these processed substrates to Layertec for the deposition of the Bragg mirror.}) and an off-the-shelf input mirror (Layertec) with a radius of curvature of \SI{5}{cm} and specified transmission $|t|^2<10^{-3}$. A cutout in the flat mirror's substrate [see Fig.~\ref{Fig:1}(b)] provides access for the scanning tip, which is a commercial scanning AFM probe (Opus 240AC-PP) glued on a metal needle. As the scanning tip interacts with the surface, we monitor the oscillation amplitude and frequency of the membrane mode, driven electrostatically at its resonance frequency with a phase-locked loop (PLL)~\cite{halg2021membrane}. The tip is coarsely displaced with piezoelectric stacks vertically along the $z$ axis (Attocube ANPz101) and horizontally in the $(x,y)$ plane (Attocube ANPx311). The fine positioning is done with an Attocube ANSxyz100 scanner. The tip position $(x, y, z)$ relative to the membrane is read out via a separate three-axis interferometer (Attocube IDS 3010) to enable hysteresis-compensated scanning and stepping over long distances.

The entire scanning microscope head, as shown in Appendix~\ref{app:cavity_design}, is placed inside a vacuum chamber at a pressure of $\sim\SI{3e-7}{mbar}$. The instrument is suspended from low spring constant springs in order to filter out external mechanical vibrations that could drive the sensor or other mechanical modes in the scanning stage or cavity apparatus. All experiments are done at room temperature.

\section{Optical and Mechanical Characterization}\label{sec:opt_mech_charac}
We now characterize the optical and mechanical properties of our cavity-detected membrane force microscope. As a first step, we extract the finesse $\mathcal{F}$ of the cavity by measuring the amplitude of the light reflected from the cavity while sweeping the wavelength of the laser, see Fig.~\ref{Fig:1}(c). We calibrate the detuning axis by sending a tone of known frequency to an EOM [not shown in Fig.~\ref{Fig:1}(a)]. The reflection dip is fitted to an asymmetric Lorentzian (i.e., Fano) lineshape \cite{Janitz2015Fabry} to extract the cavity linewidth $\kappa/2\pi = $\SI{15.9(1)}{MHz}, yielding a finesse $\mathcal{F} =\pi c/L \kappa =\SI{3.92(3)e3}{}$, with $c$ the speed of light and $L\approx \SI{2.4}{mm}$ the cavity length. Choosing a higher finesse would make the cavity harder to lock, especially inside a dilution refrigerator. For the following experiments, we lock the laser  to the cavity at a detuning of $\Delta/2\pi\approx\SI{-4.0}{MHz}$, with a measured reflected laser power $P\approx\SI{2.8}{\micro W}$.

\begin{figure}[t]
	\centering
	\includegraphics[width=1\columnwidth]{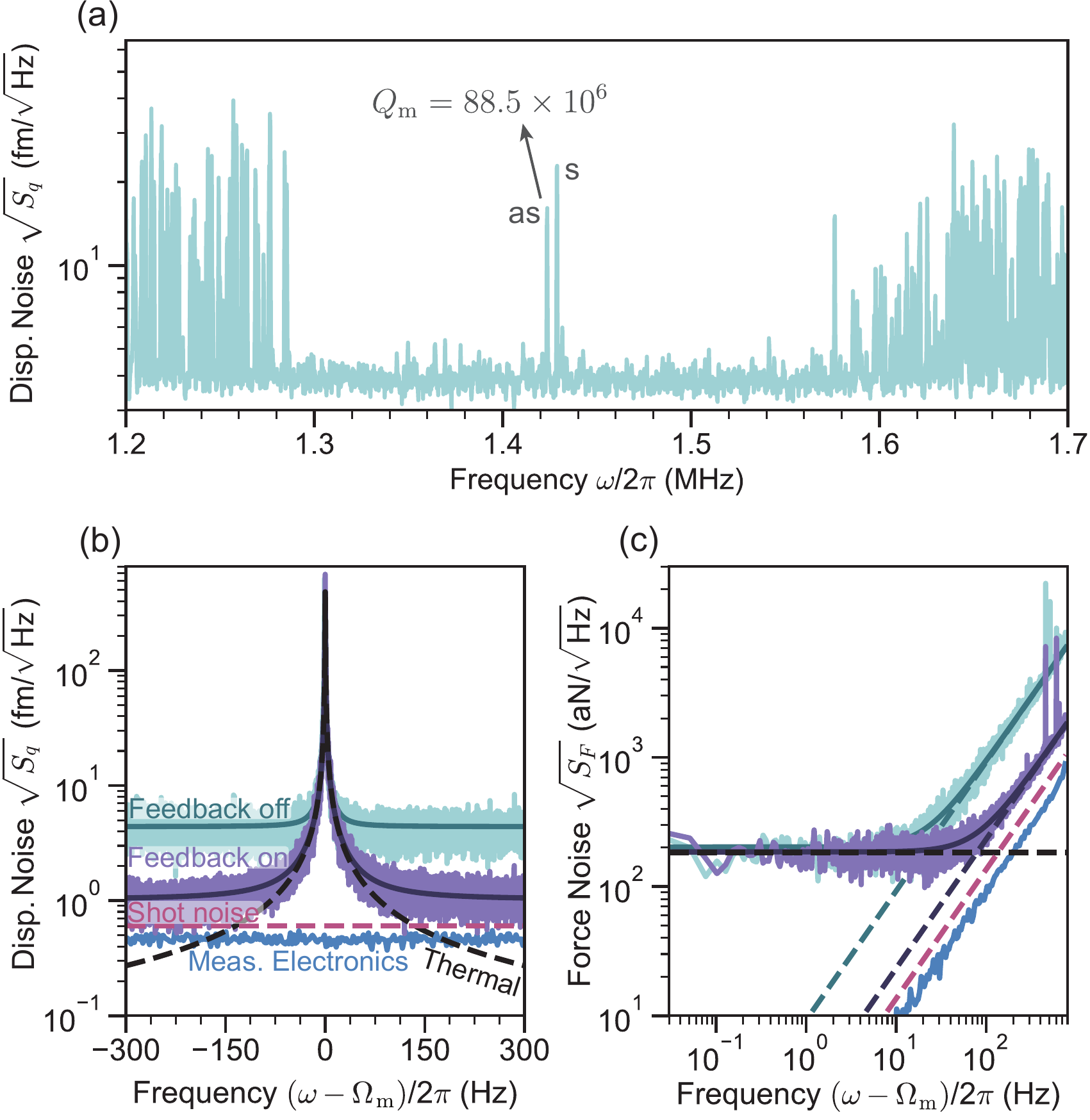}
	\caption{Mechanical properties of the membrane sensor. (a)~Displacement noise ASD measured with a laser power of $P\approx\SI{2.8}{\micro W}$ and cavity detuning $\Delta/2\pi \approx \SI{-4.0}{MHz}$. A bandgap is observed between $\omega/2\pi \approx  \SI{1.29}{MHz}$ and $\SI{1.55}{MHz}$, with two localized modes inside, the anti-symmetric (as) and symmetric (s) modes. In this work, we focus on the as mode with frequency $\Omega_\mathrm{m}/2\pi\approx\SI{1.424}{MHz}$ and quality factor $Q_\mathrm{m}= \SI{88.5(2)e6}{}$. (b)~Zoomed-in displacement noise of the anti-symmetric in-bandgap mode when the laser noise feedback is off (cyan dots) and on (purple dots). See main text for details. (c)~Force noise extracted by dividing the displacement noise ASD by $|\chi_\mathrm{eff}(\omega)|$. The force noise sensitivity is $S_{F,\mathrm{th}}^{1/2} \approx \SI{180}{aN/\sqrt{Hz}}$. The \SI{3}{dB} bandwidth of this thermally limited force noise is increased from $\approx\SI{20}{Hz}$ (cyan curve) to $\approx \SI{80}{Hz}$ (purple) by reducing the laser phase noise.}
	\label{Fig:2} 
\end{figure}

We measure our membrane's thermally driven mechanical spectrum by taking the single-sided amplitude  spectral density (ASD) of the reflected light from the cavity, as shown in Fig.~\ref{Fig:2}(a). The measured voltage ASD is converted to a mechanical displacement $q$ via the equipartition theorem $\langle q^2\rangle = k_B T_\mathrm{eff} /m_\mathrm{eff} \Omega_\mathrm{eff}^2$, with $T_\mathrm{eff} \approx \SI{13}{\kelvin}$ the effective mode temperature obtained from ringdown measurements~\footnote{The effective temperature is obtained by measuring the undamped  $\Gamma_\mathrm{m}$ (i.e., far away from the cavity resonance) and damped $\Gamma_\mathrm{eff}$ (at detuning $\Delta/2\pi \approx \SI{-4.0}{MHz}$) mechanical decay rates  assuming the mechanical mode has temperature $T=\SI{300}{\kelvin}$ when undamped, such as the damped mode has temperature $T_\mathrm{eff}= T \Gamma_\mathrm{m}/\Gamma_\mathrm{eff}$.}, $ \Omega_\mathrm{eff}$ the effective angular mechanical frequency, and $k_B$ the Boltzmann's constant. As expected, the spectrum exhibits a bandgap of mechanical modes, within which only the anti-symmetric and symmetric mechanical modes exist.

Finally, we measure the force sensitivity of our device. To do so, we optomechanically cool the anti-symmetric mode by red-detuning the laser \cite{Aspelmeyer2014Cavity}, which reduces the effective mode temperature and the effective mechanical quality factor by the same factor. As such, the cooling allows resolving the mechanical linewidth~\footnote{The mechanical linewidth $\Gamma_\mathrm{m}$ being larger implies that the the measurement time $\sim\Gamma_\mathrm{m}^{-1}$ is shorter, making the measurement less sensitive to frequency drifts.} \textit{without} affecting the thermally-limited force sensitivity~\cite{bereyhi2022perimeter}.  
We first start by measuring the position noise without any laser phase noise stabilization, but now with higher resolution and focusing on the high-$Q_\mathrm{m}$ anti-symmetric normal mode, as shown in cyan in Fig.~\ref{Fig:2}(b). Turning on the laser phase  stabilization greatly reduces the measurement noise floor, see purple data in  Fig.~\ref{Fig:2}(b). With the feedback on, the extraneous noise is on the same order of magnitude as the shot noise (red dashed line).

We fit our data with a thermally-driven displacement peak $S_{q,\mathrm{th}}^{1/2}$. We perform this fit by adding a constant  measurement imprecision noise floor $S^{1/2}_{q,\mathrm{imp}}$ added in quadrature with a thermally driven displacement peak  $S_{q,\mathrm{th}}^{1/2}(\omega) = S_{F,\mathrm{th}}^{1/2} |\chi_\mathrm{eff}(\omega)|$, where $S_{F,\mathrm{th}} = 4 m_\mathrm{eff} k_B T \Gamma_\mathrm{m}$ is the thermal force noise power spectral density (PSD) and $\chi_{\mathrm{eff}}(\omega) = [m_\mathrm{eff}(\omega^2-\Omega_{\mathrm{eff}}^2+i\omega\Gamma_{\mathrm{eff}})]^{-1}$ is the effective mechanical susceptibility. For this fit, we use the effective angular mechanical frequency $\Omega_\mathrm{eff}$, and imprecision background  $S_{q, \mathrm{imp}}^{1/2}$ as fit parameters. We use a calculated effective mass of $m_\mathrm{eff} = \SI{20}{ng}$, an effective temperature $T_\mathrm{eff} = T \Gamma_\mathrm{m}/\Gamma_{\mathrm{eff}}$, with $T = \SI{300}{K}$, as well as a bare $\Gamma_\mathrm{m}/2\pi \approx \SI{16}{\milli \hertz}$ and an effective $\Gamma_{\mathrm{eff}}/2\pi \approx \SI{340}{\milli \hertz}$ mechanical amplitude decay rates obtained from ringdown measurements. The result of these fits are shown by the cyan and purple curves in Fig.~\ref{Fig:2}(b), while the thermal contribution alone is shown as a black dashed line.  The fits yield displacement noise floors of $S_{q, \mathrm{imp}}^{1/2} \approx \SI{4.4}{\femto m/Hz^{1/2}}$ (cyan curve) and $S_{q, \mathrm{imp}}^{1/2} \approx \SI{1.0}{\femto m/Hz^{1/2}}$ (purple curve).

In Figure~\ref{Fig:2}(c), we extract the apparent force noise by dividing the displacement ASD by the mechanical susceptibility $\chi_\mathrm{eff}$. This yields a force sensitivity $S_{F,\mathrm{th}}^{1/2} \approx \SI{180}{aN/Hz^{1/2}}$ over a \SI{3}{dB} bandwidth of $\approx \SI{20}{Hz}$, which can be increased to $\approx \SI{80}{Hz}$ with the phase noise reduction scheme. That is, the laser phase stabilization results in a four-fold reduction in the displacement noise floor, and an equivalent  bandwidth increase in force noise sensitivity. 

\section{Force Scanning}

\begin{figure}[t]
	\centering
	\includegraphics[width=1\columnwidth]{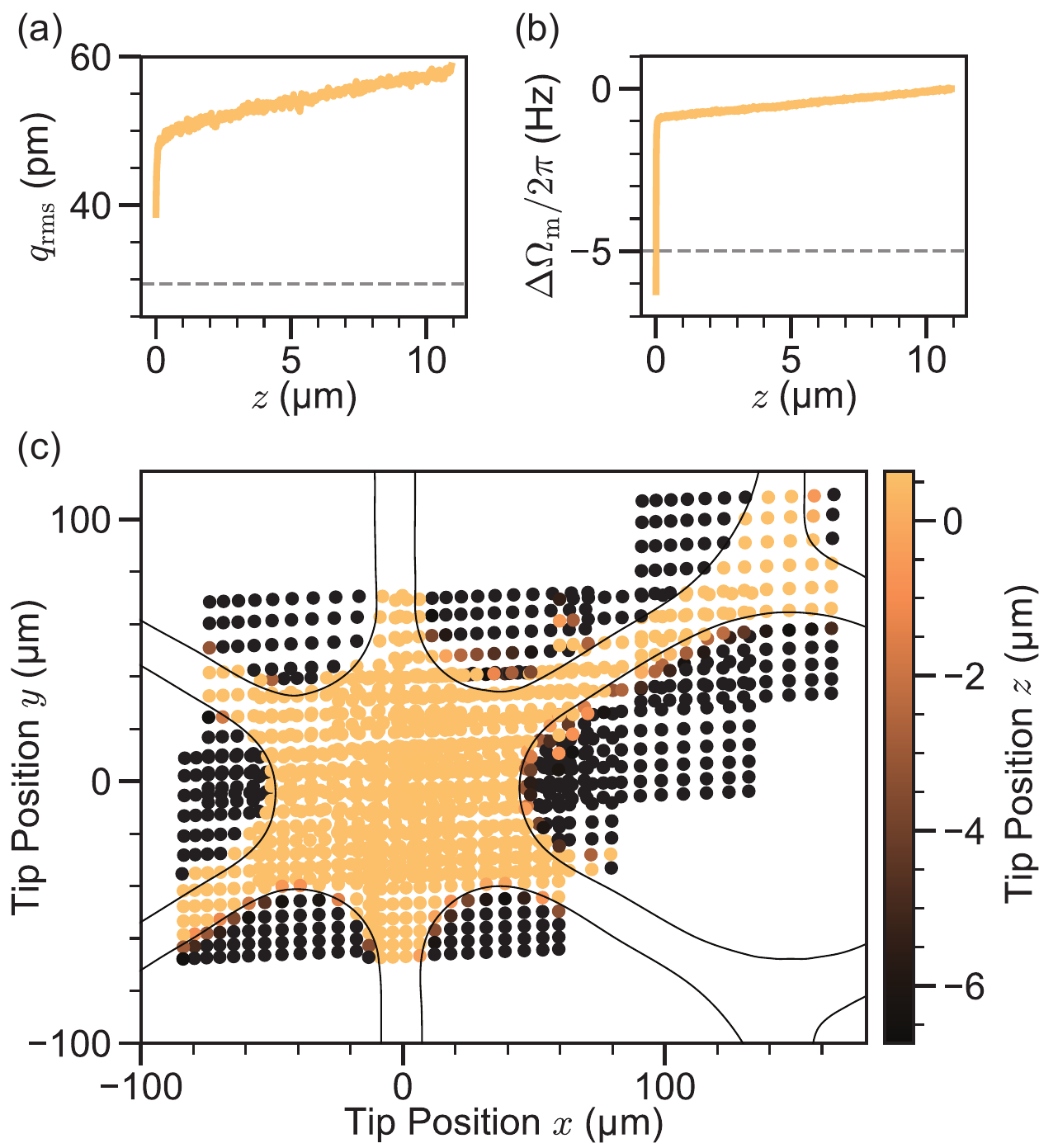}
	\caption{Touchmap image of the membrane defect. (a)~RMS amplitude $q_\mathrm{rms}$ and (b)~resonance frequency change $\Delta \Omega_\mathrm{m}$ as the tip-surface distance $z$ is reduced, at location $(x,y) =(\SI{-1}{\micro \meter}, \SI{-4}{\micro \meter})$. Sharp changes in amplitude and frequency are observed for small $z$ and are used to define a touch condition, see main text for details.  (c)~Touchmap scan of the membrane. Each point corresponds to a tip-surface approach as in (a) and (b) at a specific $(x,y)$ coordinate over the membrane. The color scale indicates the tip position when either of the touch condition (dashed line) is met. When neither criteria are fulfilled within the $z$ scanning range, the dot is colored black. The black curves indicate the membrane defect outline as extracted from a microscope image.}
	\label{Fig:3} 
\end{figure}

To demonstrate the scanning capability of the cavity setup, we record a surface topography `touchmap', shown in Fig.~\ref{Fig:3}. For a given $(x,y)$ tip position, we slowly decrease the tip-surface distance $z$ while driving the membrane electrostatically via a voltage applied to the AFM tip. The frequency of this drive is set to the membrane's resonance frequency, which we track with a phase-locked loop. At the same time, we  measure  both the membrane's displacement root-mean-square (RMS) amplitude $q_\mathrm{rms}$, shown in Fig.~\ref{Fig:3}(a), and resonance frequency change $\Delta \Omega_\mathrm{m}$, shown in Fig.~\ref{Fig:3}(b). As the tip approaches the membrane, the tip-surface electrostatic interaction results in sharp changes in both the membrane amplitude and the frequency. We register a `touch' when either the tracked membrane mode's frequency has changed by $>\SI{5}{Hz}$ or its RMS amplitude has changed by $>\SI{50}{\percent}$ of their initial values. When one of the conditions is met, we record the $z$ position of the touchpoint and retract the tip.

We repeat this touchpoint measurement over a wide range of $(x,y)$ tip positions, thus scanning the membrane surface topography and generating a touchmap~\cite{halg2021membrane} as the one shown in Fig.~\ref{Fig:3}(c). Each data point represents a tip approach, with the color indicating the tip $z$ position at which the touch criteria was fulfilled. Black dots indicate that neither touch condition was met within the available $z$ scan range. The positions of the black dots match the expected pattern of the membrane defect outline extracted from a microscope image (black curves), see Appendix~\ref{App:Sec-MembranePicture}.

In previous work, we employed this methodology to record a topography map of nanoscale objects~\cite{halg2021membrane}. However, that previous setup was limited in long-range scans by the significant hysteresis of the attocube motors in open-loop operation. Here, we implemented a three-axis interferometric stage readout (Attocube IDS 3010) to track in real time the $(x,y,z)$ position of the scanning tip relative to the membrane. The interferometer allows us to improve the precision of scans on small scales, and to stitch multiple scans into large maps. As the membrane used in this work is pristine and offers no topography objects for study, we use the outline of the central membrane defect itself  for our imaging demonstration, see Fig.~\ref{Fig:3}(c). Note that the topography transition at the edge is somewhat blurred due to the convolution between the tip and the edge of the membrane. This is due to the finite angle of the scanning tip: as the tip is lowered next to a membrane edge, this angle can result in a measured touch condition even though the tip itself is not hitting the surface. Since the tip is not perfectly symmetric, some edges of the membrane are also sharper on the touchmap.

\section{Outlook}
The work reported here takes a next step towards a force microscope capable of detecting nuclear spins on the angstrom scale. Combining the sample-on-sensor geometry of a previous membrane-based scanning force microscope~\cite{halg2021membrane} with membrane-in-the-middle cavity optomechanics~\cite{Jayich2008Dispersive, Thompson2008Strong} greatly improves the detection efficiency. For the same readout imprecision, the power reaching the membrane surface inside a cavity is a factor $\mathcal{F}$ smaller than when using a simple interferometer, cf. Appendix~\ref{app:laser_heating}, and also requires a factor $\mathcal{F}^{2}$ less input power. This leads to less optical power absorbed by the membrane and the cryostat, which in turn lowers the minimal achievable temperature~\cite{planz2023membrane}. The cavity furthermore allows dampening the motion of the sensor mode and of many out-of-bandgap mechanical modes, which otherwise could affect the measurement via e.g. thermal intermodulation noise~\cite{fedorov2020thermal, saarinen2023laser}. Further, a red-detuned cavity drive can provide effective cold damping~\cite{Schliesser2008Resolved} which, in contrast to feedback damping~\cite{Poggio2007Feedback,courty2001quantum,Rossi2018Measurement}, does not require a carefully tuned feedback loop for each mode.

In future experiments, several factors can be invoked to improve the sensitivity of the membrane mode and the overall performance of the scanning force microscope. Going to cryogenic temperatures will increase $Q_\mathrm{m}$ and at the same time reduce the thermomechanical noise. At the lowest temperature of \SI{640}{mK} demonstrated with a patterned membrane inside an optical cavity~\cite{planz2023membrane}, an ideal membrane (low mass and high-$Q_\mathrm{m}$)~\cite{saarinen2023laser} sensor mode will experience an integrated force of \SI{4e-19}{\newton} over an integration time of \SI{1}{s}, corresponding to the force generated by roughly 10 hydrogen spins in the presence of a magnetic field gradient of \SI{5}{\mega\tesla\per\meter}~\cite{mamin2012,longenecker2012high,pachlatko2024nanoscale}. This sensitivity would in principle yield sub-nanometer resolution in three dimensions for typical biological samples~\cite{Degen_2009}. For the current cavity setup, detecting such a small force will be possible within a bandwidth of \SI{20}{\hertz}, while at higher frequency offsets the remaining phase noise can become the dominant factor of the total sensor noise. Reducing our measurement noise floor to the shot noise limit, which could be achieved by cooling the cavity mirrors and/or reducing the laser phase noise with a filter cavity in series with the experiment, would increase that bandwidth to \SI{30}{\hertz}. A complete noise calculation is presented in Appendix~\ref{App:Sec-Prospects}.

Coupling nuclear spins to a membrane resonator in the MHz regime can be achieved by parametric coupling~\cite{Kosata_2020} or by engineering a near-resonance condition via dynamical back-action~\cite{visani2023near}. These detection methods will be significantly boosted when implementing Dynamic Nuclear Polarization (DNP), which has recently been demonstrated in a nanoscale experiment~\cite{tabatabaei2024large}. DNP will not only increase the signal-to-noise ratio of current protocols, but will furthermore facilitate a wide range of classical nuclear magnetic resonance protocols~\cite{slichter2013principles} beyond stochastic NanoMRI measurements~\cite{Degen_2007}.

\section*{Acknowledgements}
A. E. acknowledges financial support from Swiss National Science Foundation (SNSF) through grant 200021\_200412. A. S. acknowledges support by the European Research Council project PHOQS (Grant No. 101002179), and the Novo Nordisk Foundation (Grant Nos. NNF20OC0061866 and NNF22OC0077964). V. D. acknowledges support from the ETH Zurich Postdoctoral Fellowship Grant No.~23-1 FEL-023.

\appendix

\section{Laser heating in an optical cavity}\label{app:laser_heating}
We briefly show in this Appendix how a membrane placed inside an optical cavity requires a lower optical power to achieve the same (shot-noise-limited) imprecision noise floor as in a low-finesse interferometer. This is because a given photon measured at a photodetector has sampled the mechanical element a factor cavity finesse $\mathcal{F}$  more times than in a simple interferometer. Even though in this case the circulating power $\bar{P}_\mathrm{circ} \propto \mathcal{F} \bar{P}_\mathrm{in}$ is also a factor finesse higher than the input power $\bar{P}_\mathrm{in}$, the signal is increased by a factor $\bar{P}_\mathrm{circ} \propto \mathcal{F}$, for the same shot noise $\propto \bar{P}^{1/2}_\mathrm{in}$ at the detector. The signal-to-noise ratio is then $\mathrm{SNR} \propto \mathcal{F}\bar{P}^{1/2}_\mathrm{in}\propto \mathcal{F}^{1/2}\bar{P}^{1/2}_\mathrm{circ}$. In comparison, for a mechanical element not inside a cavity, the signal to (shot) noise ratio  increases with the squared root of the power  landing on the mechanical element, i.e. $\mathrm{SNR}\propto \bar{P}^{1/2}$, indicating that for the same power landing on the mechanical element, the imprecision noise ASD is a factor $\mathcal{F}^{1/2}$ higher than a cavity.

To verify this argument, we consider a single-sided optical cavity, where the input (power) coupling rate $\kappa_\mathrm{in}$ is equal to the cavity decay rate $\kappa$, i.e., $\kappa_\mathrm{in} = \kappa$,  for simplicity. Furthermore, we assume that the laser frequency $\omega_\mathrm{in}$ is tuned to the cavity resonance frequency $\omega_\mathrm{c}$, i.e., $\omega_\mathrm{in} = \omega_\mathrm{c}$. In this case, the displacement equivalent noise floor PSD due to shot noise is \cite{Aspelmeyer2014Cavity}
\begin{align}
    S_{q, \mathrm{imp}}(\omega) = \frac{\kappa}{8 \bar{n}_\mathrm{cav} G^2} \left[ 1 + 4\frac{\omega^2}{\kappa^2} \right],
\end{align}
with $\bar{n}_\mathrm{cav}$ the intra-cavity photon number, and $G \equiv \partial_q \omega_c$ the optomechanical coupling rate (i.e., the change in the cavity's resonance frequency per mechanical displacement $q$). 
Evaluating the imprecision force noise near the mechanical frequency, in the fast-cavity limit $\kappa \gg \Omega_\mathrm{m}$ (our relevant experimental regime), we obtain
\begin{align}
   S_{q,\mathrm{imp}} &= \frac{\kappa}{8 \bar{n}_\mathrm{cav} G^2}    \label{Eq:App-S_x1} .
\end{align}
We now express the various cavity parameters in terms of cavity finesse $\mathcal{F}$ and length $L$. For the canonical optomechanical system, where the end mirror is the mechanical element, the  optomechanical coupling rate  is $G = - \omega_\mathrm{in}/L$. The cavity decay rate can be written as $\kappa = \pi c /L\mathcal{F}$. Additionally, the mean intra-cavity circulating power $\bar{P}_\mathrm{circ}$ can be expressed in terms of  mean intra-cavity photon number $\bar{n}_\mathrm{cav}$ by noting that each photon carries energy $\hbar \omega_\mathrm{in}$ and contacts the mechanical element at each cavity round-trip time $2L/c$. Replacing these expressions in Eq.~\eqref{Eq:App-S_x1}, we obtain
\begin{align}
   S_{q,\mathrm{imp}}   = \frac{\hbar c^2 }{8  \omega_\mathrm{in} }    \frac{1}{\bar{P}_\mathrm{circ}}  \frac{\pi}{2\mathcal{F}}        \label{Eq:App-S_x_final} .
\end{align}
From Equation~\eqref{Eq:App-S_x_final}, we directly see that for the same circulating power, the imprecision PSD is reduced by a factor $\mathcal{F}$. As such, a cavity can achieve the same imprecision as a low-finesse interferometer but with a factor $\mathcal{F}$ less power impinging on the mechanical element, thus vastly reducing laser heating. For comparison, a shot-noise-limited Michelson interferometer with power $\bar{P}_\mathrm{in}$ reflecting from a (highly-reflective) membrane achieves displacement sensitivity of
\begin{align}
    S_{q, \mathrm{imp}} = \frac{\hbar c^2}{8\omega_\mathrm{in}} \frac{1}{\bar{P}_\mathrm{in}}  ,
\end{align}
which is a factor $2\mathcal{F}/\pi$ higher than a (single-port) cavity for the same laser power (i.e., same laser absorption) landing on the membrane. Note that since $\bar{P}_\mathrm{circ}\propto \mathcal{F} P_\mathrm{in}$, the cavity requires a factor $\mathcal{F}^2$ less input power.

\section{Prospects for nuclear spin sensing}\label{App:Sec-Prospects}

\begin{figure}[t]
	\centering
	\includegraphics[width=0.88\columnwidth]{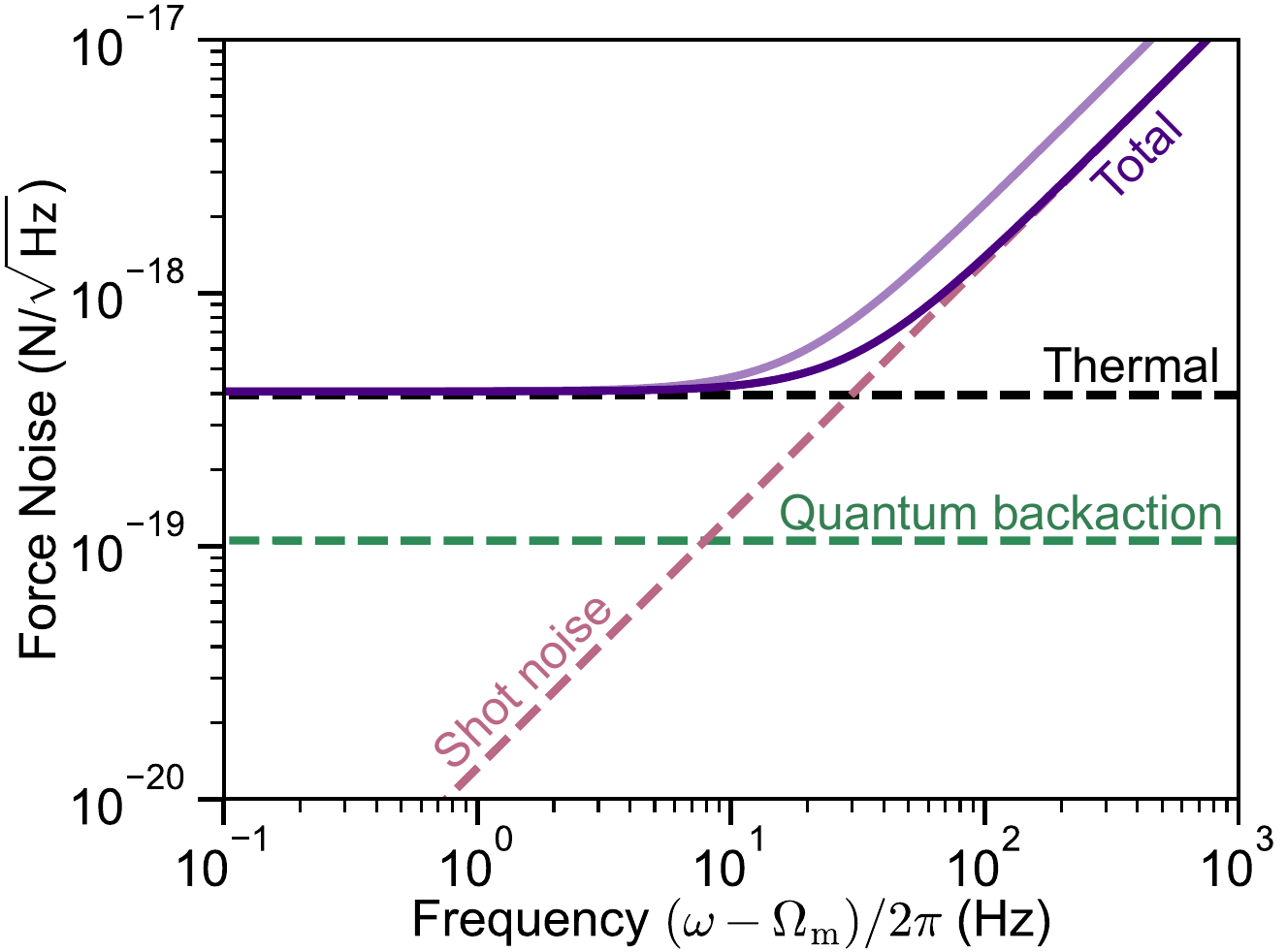}
	\caption{Force noise budget with an ideal membrane. Predicted single-sided force noise ASD as a function of frequency for a double-defect membrane with mass $m_\mathrm{eff} = \SI{0.2}{ng}$, frequency $\Omega_\mathrm{m}/2\pi = \SI{1.4}{MHz}$ and quality factor $Q_\mathrm{m} = 4 \times 10^8$ at cryogenic temperature $T = \SI{640}{mK}$, and with our current cavity. The shot noise (red dashed line), quantum backaction (green line) and thermal (black dashed line) sum up, in quadrature, to the total noise budget (purple line). We consider the current imprecision of our interferometer (light purple line), and shot-noise-limited readout (dark purple line).}
	\label{Fig:App-3} 
\end{figure}

In Fig.~\ref{Fig:App-3}, we present the calculated noise budget for our current scanning cavity with an existing membrane design optimized for force sensing. We consider a membrane thermalized at $T=\SI{640}{mK}$~\cite{planz2023membrane}, with an effective mass of $m_\mathrm{eff}=\SI{0.2}{ng}$~\cite{saarinen2023laser}  and a frequency of $\Omega_\mathrm{m}/2\pi = \SI{1.4}{MHz}$. This yields a thermal force noise of $S_\mathrm{F, \mathrm{th}}^{1/2} = \SI{0.4}{aN/Hz^{1/2}}$ (black dashed line). We also show the quantum backaction~\cite{Aspelmeyer2014Cavity} (green dashed line) for perfect quantum detection efficiency $\eta = 1$ (i.e., all photons inside the cavity get detected).  Assuming we can reach the shot noise limit (red dashed line) for our measured shot noise-equivalent displacement noise floor $S_{q,\mathrm{imp}}^{1/2} = \SI{0.6}{fm/Hz^{1/2}}$, cf. Fig.~\ref{Fig:2}(b), we would achieve a thermally limited total noise floor (dark purple line) over a bandwidth of \SI{30}{Hz}. Otherwise, if we are limited by the same displacement noise floor as presented in this work, $S_{q,\mathrm{imp}}^{1/2} = \SI{1}{fm/Hz^{1/2}}$, we will obtain a thermally limited noise floor over a \SI{18}{Hz} bandwidth.

In reality, our current measurement efficiency is most likely quite small, i.e. $\eta \ll 1$, which means that our achieved detection noise is associated with higher quantum backaction. Our measured imprecision noise already intrinsically includes the detection efficiency, so the estimated quantum backaction will be higher since there are more photons in the cavity than we actually measured. To avoid the force noise being limited by quantum backaction, the laser power should be decreased. This will come at the expense of increased measurement imprecision due to  shot noise~\cite{Rossi2018Measurement}, which will reduce the thermally limited bandwidth. Our force sensor's performance will thus be limited by an interplay between quantum backaction, setting the minimal achievable noise floor, and the  measurement imprecision, setting the bandwidth over which this noise floor is achieved.

\section{Laser frequency noise}\label{app:laser_phase_noise}
In this Appendix, we present the frequency noise measurement of our laser.

\begin{figure}[t]
	\centering
	\includegraphics[width=0.85\columnwidth]{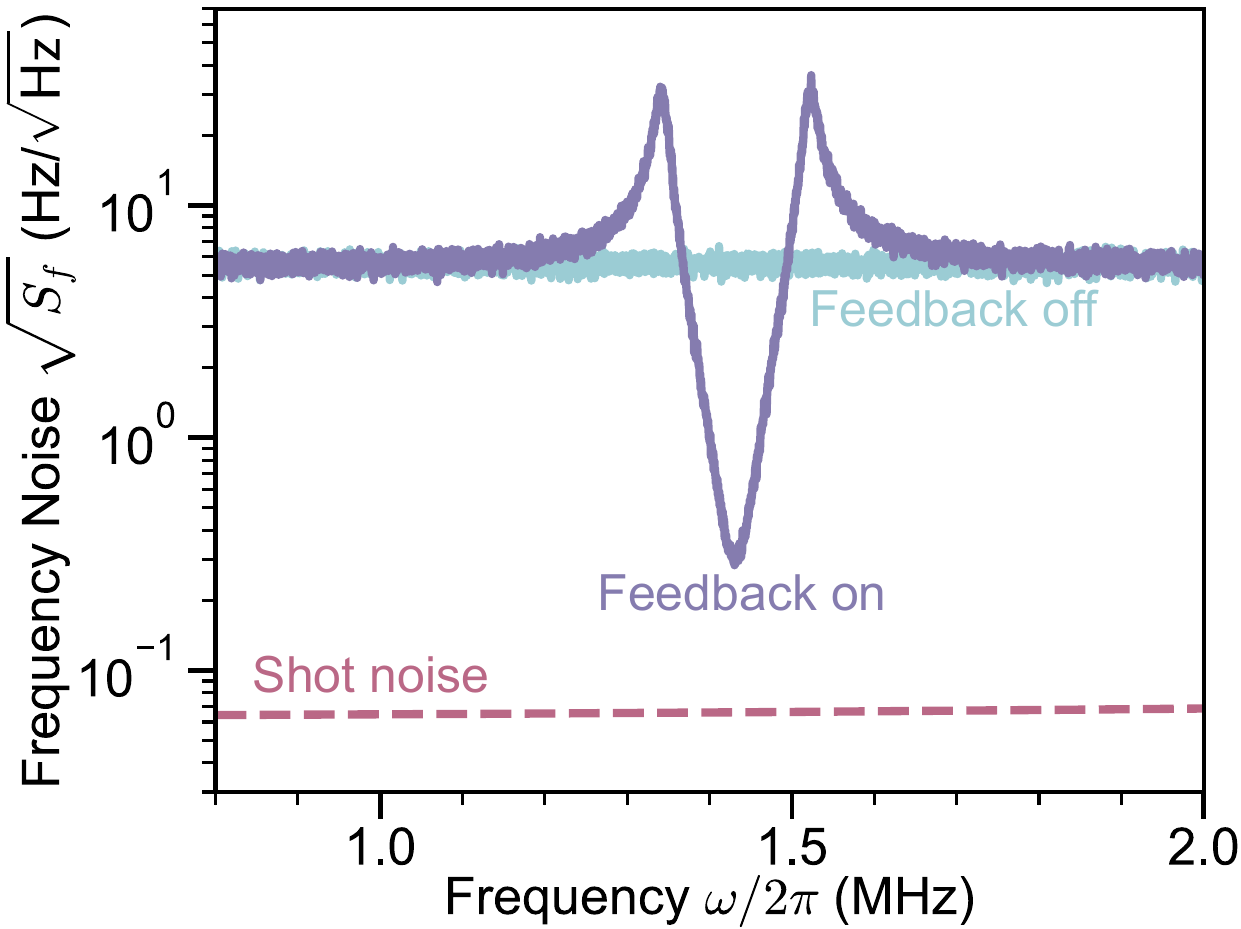}
	\caption{Laser frequency noise. Measured frequency noise ASD $S_f^{1/2}(\omega) = S_\phi^{1/2}(\omega) \omega/2\pi$ of our laser with (purple curve) and without (cyan curve) phase stabilization feedback, showing a $\sim 17$-fold reduction in frequency noise near the mechanical mode frequency $\Omega_\mathrm{m}/2\pi = \SI{1.424}{MHz}$.}
	\label{Fig:App-2} 
\end{figure}

To measure our laser phase noise after the phase stabilization loop in Fig.~\ref{Fig:1}(a) (i.e., out-of-loop), we use a self-homodyne detection scheme similar to the one described in the main text. The measured homodyne current ASD $S_I^{1/2}$ can be converted to phase noise ASD $S_\phi^{1/2}$ with \cite{parniak2021high}
\begin{align}
     \sqrt{S_\phi(\omega)}  = |I_0 D(\omega)|^{-1} \sqrt{S_I(\omega)} ,
\end{align}
where $I_0$ is the homodyne current fringe amplitude, and $D(\omega) = 1 - e^{-i\omega\tau_\mathrm{d}}$ is the interferometer delay line transfer function, with delay time $\tau_\mathrm{d}= L_\mathrm{d}/c \approx \SI{107}{ns}$ for our optical path length $L_\mathrm{d} \approx \SI{32.2}{m}$.  This measured phase noise ASD $S_\phi^{1/2}$ can be converted to frequency noise ASD $S_f^{1/2}$ via $S_f^{1/2}(\omega) = S_\phi^{1/2}(\omega)2\pi/\omega$.

In Figure~\ref{Fig:App-2}, we show the measured (out-of-loop) frequency noise of our laser with (purple) and without (cyan) phase stabilization. A reduction in the laser frequency noise by a factor $\approx 17$ can be seen near the feedback bandwidth, around frequency $\omega/2\pi = \SI{1.424}{MHz}$. Note that the shot noise and electronic noise (which is lower than the shot noise in Fig.~\ref{Fig:App-2}) are lower than the lowest achieved frequency noise of \SI{0.31}{Hz/Hz^{1/2}}, indicating that other effects (e.g. fiber noise \cite{parniak2021high}) are limiting the minimum achievable and/or detectable frequency noise.

\section{Scanning cavity}\label{app:cavity_design}
Here we present the general design of our cavity-based membrane force scanner.

\begin{figure}[t]
	\centering
	\includegraphics[width=1\columnwidth]{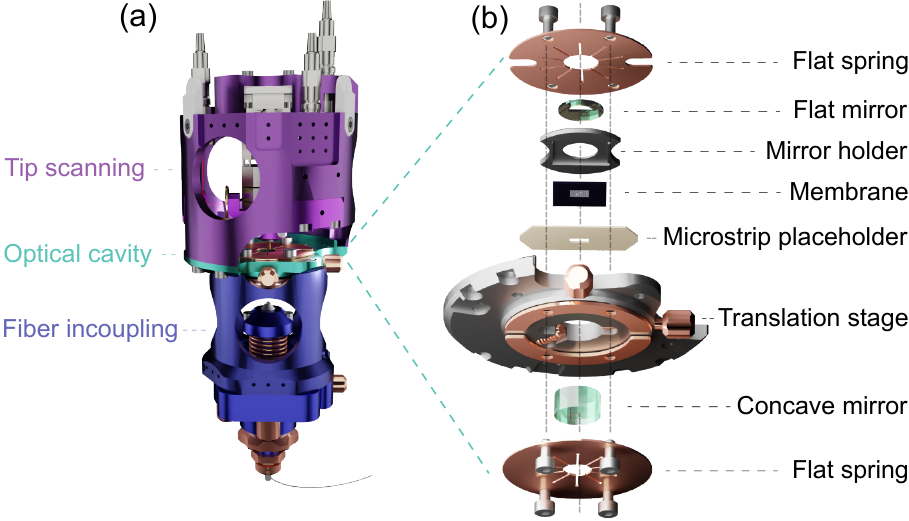}
	\caption{Scanning cavity force sensor. (a) False color rendering of the three different stages of the scanning cavity. The AFM tip scanning stage (purple, top), the membrane-in-the-middle cavity (cyan, middle) and the fiber incoupling apparatus (blue, bottom).  The fiber incoupling stage allows for an optical fiber (here, terminated with a GRIN lens, to focus the light), to be mode matched to the cavity via screws-controlled in-plane translation and tilt. (b) Exploded view of the membrane-in-the-middle cavity. Two flat springs softly clamp the stack in position. The translation stage allows for in-plane displacement of the concave mirror. The microstrip placeholder acts as a placeholder for future chips to deliver radiowave pulses to nuclear spin samples.}
	\label{Fig:App-1} 
\end{figure}

A false color rendering of the scanning cavity force sensor is shown in Fig.~\ref{Fig:App-1}(a). There are three main parts in the scanning cavity: (1)~a fiber incoupling stage, shown in blue, where an optical fiber with a gradient index (GRIN) lens at its end sends focused light to the cavity (and collects the reflected light), and can be aligned to the cavity with tilt and translations screws, (2)~the optical membrane-in-the-middle cavity, shown in cyan, where the (curved) input mirror of the cavity can be translated via two translation screws to align the incoming light to the membrane's sensing defect, and (3)~the tip scanning stage, shown in purple, that allows for 3-axis fine and coarse positioning of an AFM tip, along with an interferometric readout of the tip position. 

The various parts of the membrane-in-the-middle cavity are shown in Fig.~\ref{Fig:App-1}(b). Our double-defect membrane is placed on a microstrip place holder (which will be replaced with a microstrip chip of the same height in future experiments, to perform magnetic resonance force microscopy), while a mirror holder fixes the flat mirror in place. The concave input mirror can be displaced in the $(x,y)$ plane with the translation stage to align the cavity mode to the `readout' membrane defect. The full membrane-in-the-middle cavity stack is clamped with a flat spring at both of its ends.

Our parts are made of non-magnetic materials since we are planning to cool down this instrument and perform magnetic force resonance microscopy, requiring large magnetic fields. Most parts are made of either titanium or beryllium copper (BeCu$_2$). We use two different materials to avoid cold welding.

\section{Double-defect membrane}\label{App:Sec-MembranePicture}

\begin{figure}[t]
	\centering
	\includegraphics[width=0.75\columnwidth]{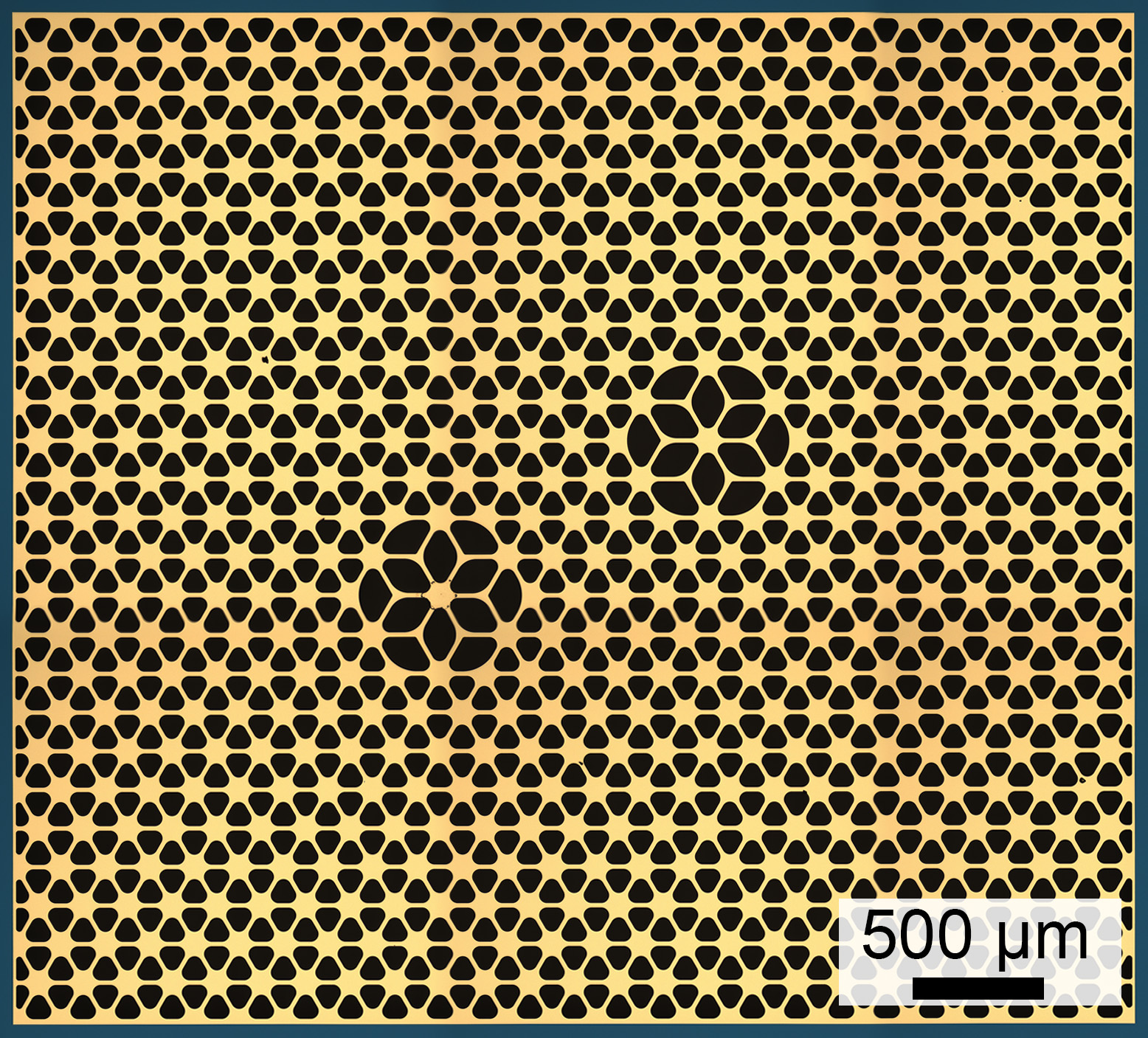}
	\caption{Microscope picture of our double-defect membrane. The membrane (yellow) is made of suspended 20-nm-thick high-stress silicon nitride (Si$_3$N$_4$), which is held by the silicon frame (blue). }
	\label{Fig:App-4} 
\end{figure}

In Figure~\ref{Fig:App-4}, we show a picture of our double-defect membrane. The membrane (yellow) is patterned with a phononic crystal to both soft-clamp the membrane and prevent radiation of energy through the substrate (blue)~\cite{Tsaturyan2017Ultracoherent}. This allows for high quality factors for the bandgap modes. The two defects that can be seen in the photograph have out-of-plane motion and are coupled due to evanescent overlap. As such, they hybridize into symmetric and anti-symmetric modes~\cite{halg2021membrane, catalini2020soft}.

In this work, we image one of the defects, the `sensing' defect, c.f. Fig.~\ref{Fig:3}, while using the other defect as a `readout' defect which we readout with our laser.

\addcontentsline{toc}{section}{References}
\bibliographystyle{bibstyle-jack}
\bibliography{errthing.bib}

\end{document}